\documentclass[journal, twoside]{IEEEtran}
\usepackage{indentfirst}
\usepackage{graphicx}
\usepackage{amsmath}
\usepackage{amssymb}
\usepackage{amsfonts}
\usepackage{mathrsfs}
\usepackage{leftidx}
\usepackage{color}
\usepackage{amsmath}
\usepackage{arydshln}
\usepackage{amsthm}
\usepackage{ragged2e}
\usepackage{cite}
\usepackage{enumerate}
\usepackage{longtable}
\usepackage{float}
\usepackage{stfloats}
\usepackage{hyperref}
\usepackage{algpseudocode}
\usepackage{algorithm}
\usepackage[caption=false,font=footnotesize]{subfig}
\usepackage{tabularx}
\usepackage{makecell}
\usepackage{url}
\usepackage{tabularx}

\usepackage{multirow} 
\usepackage{booktabs}
\theoremstyle{plain}

\usepackage{caption}
\usepackage[table]{xcolor}
\usepackage{multirow}
\usepackage{array}

\newcolumntype{P}[1]{>{\raggedright\arraybackslash\footnotesize}m{#1}}
\newcolumntype{A}[1]{>{\centering\arraybackslash\footnotesize}m{#1}}

\usepackage[table,usenames,dvipsnames]{xcolor}
\definecolor{aa}{RGB}{175,238,238}
\definecolor{bb}{RGB}{255,255,255}
\usepackage{bm}
\usepackage{makecell}

\begin{document}

\title{Towards Semantic Internet of Everything in the Age of Agentic AI}

\author{Dayu Fan, Rui Meng,~\IEEEmembership{Member,~IEEE,} Yunfei Liu, Shuai Ai,
Xiaodong Xu,~\IEEEmembership{Senior Member,~IEEE,}

Yiming Liu,~\IEEEmembership{Member,~IEEE,} Huishi Song, Han Meng, Lexi Xu,~\IEEEmembership{Senior Member,~IEEE,}

and
Ping Zhang,~\IEEEmembership{Fellow,~IEEE}

\thanks{
This work was supported in part by the National Key R\&D Program of China under Grant 2025YFF0514703; in part by the National Natural Science Foundation of China under Grant 62501066; in part by the S\&T Program of Hebei under Grant 262X0405D; and in part by the China Postdoctoral Science Foundation under Grant 2026M793832. 
\textit{(Corresponding author: Rui Meng.)}

Dayu Fan is with the State Key Laboratory of Networking and Switching Technology, Beijing University of Posts and Telecommunications, Beijing 100876, China, and also with 
the ZGC Institute of Ubiquitous-X Innovation and Applications, Beijing 100083, China (e-mail: fandayu@bupt.edu.cn).

Rui Meng and Xiaodong Xu are the State Key Laboratory of Networking and Switching Technology, Beijing University of Posts and Telecommunications, Beijing 100876, China, and also with Xiong'an Aerospace Information Research Institute, Xiong'an 070001, China (e-mail: buptmengrui@bupt.edu.cn; xuxiaodong@bupt.edu.cn).

Yunfei Liu, Shuai Ai, Yiming Liu, and Ping Zhang are the State Key Laboratory of Networking and Switching Technology, Beijing University of Posts and Telecommunications, Beijing 100876, China (e-mail: feifeily712@gmail.com; asasas@bupt.edu.cn; liuyiming@bupt.edu.cn; pzhang@bupt.edu.cn).

Han Meng is with the Institute of Network and IT Technology, China Mobile Research Institute, Beijing 100053, China (email: menghanyjy@chinamobile.com).

Huishi Song is with the ZGC Institute of Ubiquitous-X Innovation and Applications, Beijing 100083, China (email: songhuishi@zgc-xnet.com).

Lexi Xu is with the Research Institute, China United Network Communications Corporation, Beijing, China (e-mail: davidlexi@hotmail.com).

}}

\maketitle

\begin{abstract}
Semantic communication improves task effectiveness by transmitting task-relevant information. However, most existing schemes remain organized as task-specific, end-to-end pipelines, which are difficult to reuse across models, applications, and deployment environments. Against this background, we propose the Semantic Internet of Everything (SIoE), a composable service architecture that represents heterogeneous communication and artificial intelligence (AI) functions as capability-profiled services and coordinates them according to application objectives. SIoE comprises three planes: a task and service plane, an agentic orchestration plane, and a semantic capability plane. In this framework, task requirements are captured via a semantic service-level agreement (SLA), while an agentic planner discovers and composes candidate capabilities under deterministic compatibility, resource, privacy, and policy validation. Feedback from the communication, semantic, and task levels enables continuous adaptation and replanning. A lightweight vehicle-to-everything case study illustrates profile-grounded capability planning under explicit service constraints. The results demonstrate the feasibility of decoupling service objectives from fixed communication implementations and also highlight key open challenges, including semantic SLA design, capability interoperability, scalable planning, and trustworthy execution.
\end{abstract}


\section{Introduction}

Bit-oriented communication architectures and conventional quality-of-service (QoS) mechanisms have provided a reliable foundation for the Internet of Things (IoT) and large-scale 5G deployments. As the Internet of Everything (IoE) evolves toward increasingly intelligent and autonomous services, however, communication systems must support not only data delivery but also distributed perception, reasoning, decision making, and action. Existing network abstractions lack explicit representations of task intent, semantic relevance, or the capabilities and limitations of distributed AI models. 

Semantic communication addresses this limitation by prioritizing task-relevant information and downstream effectiveness over faithful source-bit recovery. Existing studies\cite{nguyen2025contemporary} span semantic encoding, joint source-channel coding, structured representations, generative reconstruction, and task-oriented transmission across modalities and applications. Nevertheless, most existing schemes remain fixed end-to-end pipelines, tailored to specific tasks, modalities, channel assumptions, or encoder-decoder pairs. Consequently, changes in objectives, network conditions, deployment scenarios, or computing resources frequently necessitate manual reconfiguration, retraining, or replacement.

The resulting challenge is therefore broader than the design of any individual semantic codec: heterogeneous communication and artificial intelligence (AI) functions must be described, discovered, validated, and reused across applications. Recent modular semantic communication studies have begun to expose previously monolithic processing stages as explicit and replaceable skills \cite{fu2026skillcom,meng2026skillcomm}. However, modularizing one transmission workflow does not by itself provide a service architecture for coordinating capabilities from different model families, deployment domains, and communication layers.

Agentic AI provides a complementary mechanism for addressing this. An AI agent can interpret  high-level objectives, invoke specialized tools, observe execution results, and revise subsequent actions through a closed-loop workflow\cite{acharya2025agentic}. Recent wireless agent frameworks have demonstrated executable workflow construction, tool-augmented reasoning, and interactive network management \cite{tong2026wirelessagent++,chen20266gagentgym,li2026agentic}. Related studies on semantic agent communication show that user intention, receiver feedback, and multimodal task context can guide content selection and semantic refinement \cite{liu2025wireless,jiang2026intention,jiang2026agentcomm}. These developments suggest that Agentic AI is most useful not as a replacement for specialized communication and task-specific models, but rather as a coordination mechanism that aligns them with application objectives and runtime observations.

Driven by this insight, we propose the Semantic Internet of Everything (SIoE), a composable service architecture for the task-driven coordination of heterogeneous communication and AI capabilities. A capability may implement semantic extraction, source or channel coding, signal processing, reconstruction, knowledge access, task inference, or evaluation. Beyond a callable interface, each capability exposes a machine-readable profile that details its input and output types, operating conditions, compatibility dependencies, deployment requirements, resource costs, and expected task-level performance.

Unlike existing frameworks that center on agent interaction, network-management workflows, or modular stages within a transmission process, SIoE models heterogeneous communication and AI functions as capability-profiled services and composes them into validated information flows. Crucially, pipeline construction is governed by typed interfaces, model pairings, deployment policies, and measured performance, rather than relying solely on natural-language tool selection.

Application requirements are captured in a semantic service-level agreement (SLA). Alongside conventional constraints such as latency, bandwidth, reliability, and energy consumption, the semantic SLA may also specify the required task output, minimum task quality, acceptable semantic distortion, privacy restrictions, and deployment preferences. Given these requirements, an orchestration agent discovers candidate capabilities, constructs a pipeline, and adapts its configuration based on feedback from the communication, semantic, and task levels. Deterministic compatibility and policy mechanisms validate the proposed pipeline prior to execution.

The envisioned SIoE is structured into three logical planes: a task and service plane that captures application intent and semantic SLA requirements, an agentic orchestration plane responsible for capability discovery, composition, validation, execution, and adaptation, and a semantic capability plane encompassing communication, semantic processing, knowledge, and downstream task functions. These planes describe system responsibilities rather than replacements for the conventional protocol stack, and their components may be distributed across devices, edge servers, and cloud platforms.
The main contributions of this article are summarized as follows:

\begin{itemize}
\item We identify an emerging transition from task-specific semantic transceivers toward reusable communication and AI capabilities, and clarify the role of Agentic AI as a constrained orchestration mechanism rather than a replacement for specialized models.
\item We develop a three-plane SIoE architecture that translates semantic SLA requirements into validated capability pipelines with feedback-driven adaptation across device, edge, and cloud resources.
\item We demonstrate SIoE with a lightweight V2X prototype for SLA-constrained capability selection and identify challenges in capability specification, interoperability, orchestration, and trustworthy execution.

\end{itemize}

\section{From Semantic Communications to SIoE}

Semantic communication improves how task-relevant information is represented and transmitted, whereas Agentic AI translates high-level objectives into executable and adaptive workflows. SIoE connects these technologies by treating communication and AI functions as reusable capabilities, which can be selected and coordinated according to objective requirements.

\subsection{Task-Oriented Semantic Communication}

Conventional communication systems primarily optimize for reliable bit delivery subject to constraints such as throughput, latency, reliability, and energy efficiency. Semantic communication extends this objective by assessing whether the received information remains useful for a downstream task. Depending on the application, the transmitted representation may constitute reconstructed source content, semantic features, structured messages, selected regions of interest, or latent representations consumed directly by a task model. Accordingly, evaluation criteria have expanded beyond bit error rate and source distortion to encompass semantic fidelity and task performance. Recent intention- and agent-oriented methods prioritize or protect information according to its contribution to the intended task \cite{jiang2026intention,jiang2026agentcomm}, demonstrating that transmission policies can be guided by application objectives.

Nevertheless, most systems still rigidly bind semantic representations, encoders, decoders, channel assumptions, and downstream models to a specific task. SkillCom and SkillComm enhance flexibility by exposing processing stages as replaceable skills for semantic abstraction, transmission, repair, and sequential execution \cite{fu2026skillcom,meng2026skillcomm}. However, this modularity remains confined to a predefined semantic communication workflow rather than enabling capabilities to be reused and composed across tasks and deployments.

\subsection{Agentic Orchestration and Capability Composition}
Recent studies on semantic-based agent communication examine semantic exchange, reasoning, and collaboration among intelligent agents \cite{zhang2026towards}, while wireless Agentic AI frameworks employ planning and executable tools for network management. WirelessAgent++ organizes wireless problem-solving into modular workflows, and 6GAgentGym provides typed tools and observable environments \cite{tong2026wirelessagent++,chen20266gagentgym}. Related work has also explored intent-aware physical-layer intelligence and tool-augmented network planning \cite{li2026agentic,zhang2026small,zhang2024satelliteagent}. These studies advance agent interaction and network control, yet leave open the question of how the underlying communication and AI functions can be described and composed into validated end-to-end pipelines.

SIoE tackles this complementary challenge by exposing semantic extraction, coding, transmission, reconstruction, inference, and evaluation functions as discoverable services. Safe composition demands more than mere function names: a planner must be aware of their input and output types, model pairings, operating and deployment conditions, resource demands, and expected task performance. Accordingly, SIoE augments invocation interfaces with machine-readable capability profiles. It constructs pipelines from task requirements, validates compatibility and policy constraints, and adapts execution based on runtime feedback.
SIoE complements, rather than replaces, conventional radio access, routing, transport, security, and QoS mechanisms. It introduces a task-oriented service abstraction for coordinating communication and AI capabilities according to application outcomes. Table~\ref{tab:sioe_evolution} summarizes the progression from connectivity-oriented IoT, through task-specific semantic communication, to capability-oriented SIoE.

\begin{table*}[htbp]
\centering
\caption{Evolution from Connectivity-Oriented IoT to SIoE}
\label{tab:sioe_evolution}
\footnotesize
\setlength{\tabcolsep}{0.45em}
\renewcommand{\arraystretch}{1.15}
\resizebox{\linewidth}{!}{%
\begin{tabular}{p{2.5cm}p{4.4cm}p{4.8cm}p{5.5cm}}
\toprule
\textbf{Dimension}
&
\textbf{Conventional IoT and Networked AI}
&
\textbf{Semantic Communication}
&
\textbf{SIoE}
\tabularnewline
\midrule

Service objective
&
Reliable data delivery between connected devices
&
Delivery of information needed by a predefined downstream task
&
Composition of communication and AI services to meet requirements expressed by a semantic SLA
\tabularnewline
\midrule

System organization
&
Protocol functions coupled with application-specific AI
&
A fixed semantic encoder-channel-decoder pipeline for one task
&
Reusable communication and AI capabilities composed into validated pipelines
\tabularnewline
\midrule

Adaptation and evaluation
&
QoS adjusted using link state and traffic conditions; evaluated by throughput, latency, reliability, and energy
&
Representation and coding adapted to channel conditions for a fixed task \newline Evaluated by semantic fidelity and task accuracy
&
Capabilities selected or replaced using the semantic SLA, availability, and runtime feedback \newline Evaluated by task quality, cost, latency, privacy, and reliability
\tabularnewline
\midrule

Role of AI
&
Application inference or optimization of individual network functions
&
Task-specific semantic encoding, reconstruction, and inference
&
Agentic planning proposes pipelines, while typed interfaces, model pairings, and policies validate execution
\tabularnewline
\bottomrule
\end{tabular}%
}
\end{table*}

\section{A Three-Plane Reference Framework for Composable SIoE}
\label{sec:framework}

\begin{figure*}[t]
    \centering
    \includegraphics[width=\textwidth]{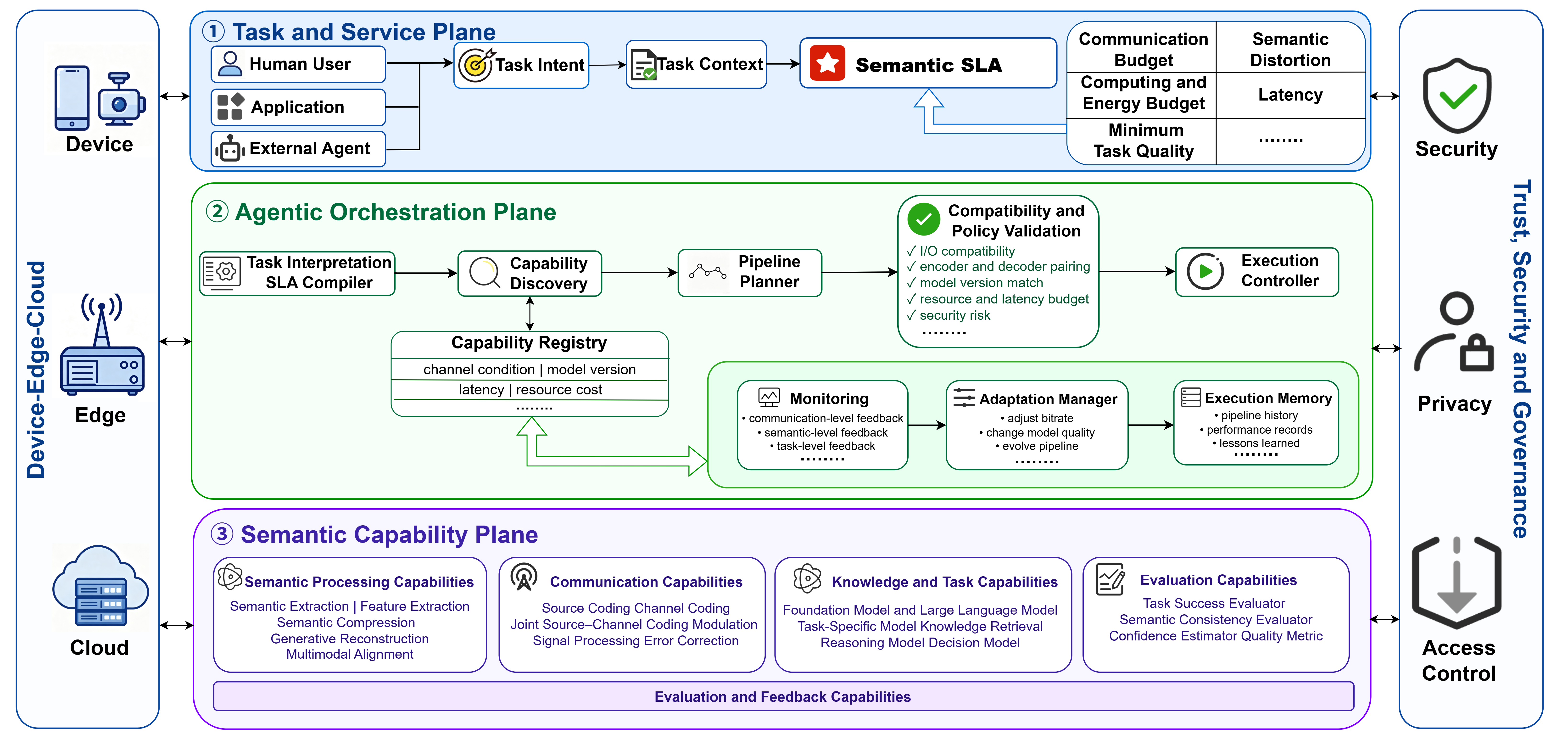}
\caption{The proposed three-plane SIoE architecture, where the task and service plane converts task intent and context into a semantic SLA; the agentic orchestration plane discovers, composes, validates, executes, and adapts capability pipelines using multi-level feedback; and the semantic capability plane provides reusable communication, semantic, knowledge, task, and evaluation functions. The framework spans device, edge, and cloud resources under trust, security, privacy, and access-control constraints.}
\label{fig:sioe_framework}
\end{figure*}

SIoE organizes task-driven communication and AI services through the three-plane framework shown in Fig.~\ref{fig:sioe_framework}. The task and service plane describes the intended application outcome and service requirements. The agentic orchestration plane translates these requirements into a validated and executable capability pipeline. The semantic capability plane provides the communication, semantic processing, knowledge, task, and evaluation functions required by the pipeline. The three planes represent logical responsibilities rather than physical protocol layers. They operate above and across conventional communication infrastructures, while radio access, routing, transport, security, and QoS mechanisms continue to provide connectivity and resource support. Components from different planes may be deployed on devices, edge servers, or cloud platforms according to latency, privacy, computing, and reliability requirements.

The framework follows four principles. First, an application specifies the required outcome without selecting a particular codec, model, or deployment location. Second, communication and AI functions are exposed as reusable capabilities with executable interfaces and machine-readable profiles. Third, agentic planning is constrained by deterministic compatibility and policy validation before execution. Finally, communication-level, semantic-level, and task-level observations are used to adapt the selected pipeline.

\subsection{Task and Service Plane}

The task and service plane serves as the entry point for a human user, an application, or another agent. It specifies what the system should accomplish, without requiring the requester to understand the internal implementation of the available communication and AI models. A request may describe the input modality, task context, expected output, priority, and service preferences. For example, a vehicle may request the identification and location of a road hazard together with an appropriate driving response.

These requirements are represented by a semantic SLA, denoted by $\mathcal{S}_{\tau}=\{\mathcal{Y}_{\tau},\mathcal{R}_{\tau}\}$, where $\mathcal{Y}_{\tau}$ specifies the required task output and $\mathcal{R}_{\tau}$ captures quality, semantic fidelity, latency, resource, and policy requirements. The relative importance of these requirements varies by task. Reconstruction may emphasize semantic fidelity, whereas control may prioritize a reliable decision before a deadline. The semantic SLA also distinguishes hard constraints from soft preferences.

Table~\ref{tab:sla_matching_example} provides a qualitative V2X example. The entries illustrate how task requirements constrain capability selection rather than prescribing universal deployment thresholds. Experiment-specific numerical requirements are introduced separately in the case study. The example shows that capability discovery cannot rely only on function names. Two capabilities may both provide image transmission or object recognition while differing in their output schemas, operating regions, deployment restrictions, and task performance. The task and service plane therefore provides the structured requirements against which candidate capabilities are retrieved and evaluated.

\begin{table*}[t]
\centering
\caption{Qualitative Example of Semantic SLA-to-Capability Matching}
\label{tab:sla_matching_example}
\footnotesize
\setlength{\tabcolsep}{0.5em}
\renewcommand{\arraystretch}{1.15}
\resizebox{\linewidth}{!}{%
\begin{tabular}{p{2.5cm}p{6.7cm}p{6.4cm}}
\toprule
\textbf{Requirement}
&
\textbf{Illustrative V2X Request}
&
\textbf{Capability Information Needed for Matching}
\\
\midrule

Task output
&
Road-hazard category, spatial location, confidence, and recommended response
&
Output schema and task functions that can produce the requested information
\\
\midrule

Task quality and semantic preservation
&
Retain information required to recognize safety-relevant objects and support a reliable decision
&
Profiled task performance, semantic-consistency measurements, and returned confidence
\\
\midrule

Latency and resource preference
&
Complete the task within its service deadline while avoiding unnecessary communication and computing cost
&
Measured or estimated latency, communication overhead, computing demand, and deployment location
\\
\midrule

Privacy and deployment
&
Keep sensitive raw sensor data on the vehicle or within a trusted edge domain
&
Data-export policy, supported execution environment, and privacy attributes
\\
\midrule

Compatibility
&
Connect semantic extraction, transmission, reconstruction, and reasoning functions without representation mismatch
&
Input and output schemas, model-pairing relationships, preprocessing requirements, and version dependencies
\\
\bottomrule
\end{tabular}%
}
\end{table*}

\subsection{Agentic Orchestration Plane}

The agentic orchestration plane translates the task profile and semantic SLA into an executable communication workflow. Its role is to coordinate specialized capabilities, rather than to directly perform semantic coding, signal processing, or task inference.

The process begins with task interpretation and SLA compilation. Natural-language requests, application messages, or agent-generated objectives are converted into a structured task profile. Hard constraints and soft preferences are separated, and the information required for capability retrieval is identified.

A capability registry maintains the available functions together with their interfaces, deployment locations, dependencies, operating conditions, and measured or estimated performance. Capability discovery retrieves candidates that may satisfy the request, after which a pipeline planner organizes them into a dependency-aware workflow. A pipeline may contain semantic extraction, source or channel coding, transmission, reconstruction, task inference, and evaluation capabilities. Its configuration may also specify coding rate, model quality, deployment location, or channel parameters.

A candidate pipeline proposed by an agent is not executed immediately. Compatibility validation checks data schemas, semantic units, tensor or symbol formats, model versions, and required encoder-decoder pairings. Policy validation checks latency, resource, deployment, privacy, access-control, and safety constraints. Candidates that violate hard requirements are rejected before execution. After validation, the execution controller invokes the selected capabilities and manages the transfer of intermediate results. Monitoring functions collect communication-level observations, such as channel quality and decoding status; semantic-level observations, such as reconstruction consistency and semantic confidence; and task-level observations, such as inference confidence and task completion.

The adaptation manager compares these observations with the semantic SLA. Parameter adaptation may change the coding rate, model quality, or retransmission policy without altering the pipeline structure. Structural adaptation may replace a capability, migrate a model, add an auxiliary semantic branch, or select another prevalidated pipeline. Execution memory records successful configurations and observed performance so that subsequent decisions can use both registered profiles and previous execution evidence.

The orchestration loop can therefore be summarized as
\(\textbf{Discover}\rightarrow\textbf{Compose}\rightarrow\textbf{Validate}\rightarrow\textbf{Execute}\rightarrow\textbf{Adapt}\).
Agentic reasoning provides flexibility in interpreting tasks and proposing workflows, while deterministic mechanisms enforce compatibility and service constraints.

\subsection{Semantic Capability Plane}

The semantic capability plane encompasses the functions that perform the actual communication, semantic, knowledge, and task processing. As illustrated in Fig.~\ref{fig:sioe_framework}, these functions include semantic extraction and compression, source and channel coding, modulation and signal processing, generative reconstruction, knowledge retrieval, foundation or task-specific models, reasoning, decision making, and task evaluation.

A capability may be implemented by a learned model, a conventional communication algorithm, a knowledge service, or a combination of these components. Each capability provides an invocation interface and a capability profile. The invocation interface specifies how the function is called, including its inputs, outputs, parameters, status information, and failure responses. The capability profile describes whether and under which conditions the function is suitable for a particular pipeline.

The profile records four categories of information. Functional information describes the supported operation, modality, input and output schemas, semantic-unit type, and returned confidence. Operating information identifies supported channel conditions, source formats, resolutions, waveform assumptions, and required side information. Performance and resource information describes expected task quality, semantic fidelity, latency, memory, computation, communication overhead, and energy consumption. Dependency and governance information records model pairings, preprocessing requirements, software versions, deployment environments, provenance, access policies, and trust attributes.

The distinction between an invocation interface and a capability profile is important. A software wrapper may explain how to execute a model without indicating whether the model is compatible with adjacent components or whether it satisfies the current operating and service requirements. In particular, semantic encoders and decoders cannot be connected solely because their function names appear compatible. Their latent representations, checkpoints, normalization procedures, symbol formats, and channel assumptions must also match.

Evaluation functions are treated as capabilities rather than external afterthoughts. Task-success evaluators, semantic-consistency metrics, confidence estimators, and quality metrics return the evidence required by the orchestration plane. Provenance and integrity information is also needed because incorrect capability descriptions or manipulated feedback may result in invalid or unsafe adaptation decisions.

\subsection{Cross-Plane Operation and Deployment}

Cross-plane operation begins when a requester submits an objective to the task and service plane. The objective is compiled into a task profile and semantic SLA. The orchestration plane then retrieves compatible capabilities, constructs and validates candidate pipelines, and selects an executable configuration. The capability plane performs the required operations across available device, edge, and cloud resources. Execution measurements and task outcomes are returned to the orchestration plane, which retains, adjusts, or replaces the pipeline according to the semantic SLA.

This separation establishes clear control boundaries. The task and service plane defines the requested outcome but does not select individual models. The orchestration plane coordinates capabilities but does not replace numerical communication or AI processing. The capability plane performs registered functions without independently reinterpreting the original task objective.

Deployment is determined by service constraints. Devices may host sensors, lightweight semantic extractors, codecs, and local task models for privacy-sensitive or latency-critical processing. Edge platforms may provide regional registries, larger models, compatibility validation, and low-latency adaptation. Cloud platforms may maintain global capability repositories, foundation models, long-term execution histories, and model-management services.

The framework also separates adaptation timescales. Fast adaptation changes parameters or selects among prevalidated local pipelines according to current communication and task feedback. Slower processes update capability profiles, deploy new model versions, synchronize execution histories, and revise orchestration policies. Trust, security, privacy, and access control apply across all three planes and constrain capability registration, discovery, execution, and feedback.

\section{V2X Case Study: Extensible Agentic Planning over Profiled Capabilities}
\label{sec:case_study}

This section presents a lightweight V2X instantiation of the SIoE framework. The case study examines how an agentic orchestrator can plan over heterogeneous communication capabilities using measurable operating evidence, while deterministic mechanisms enforce compatibility and semantic SLA constraints. The planning space is extensible: new transmission, semantic-processing, and evaluation capabilities can be registered without changing the service request. For reproducibility, the prototype uses two transmission capabilities whose performance is measured offline and stored in their profiles. At runtime, the planner compares these profiles using an explicit SLA-based rule. The offline measurements provide reusable evidence rather than defining a fixed workflow.

\subsection{Service Request and Agentic Planning Space}

We consider a vehicle that transmits a road-scene image to a nearby edge server for perception-oriented processing. The task and service plane specifies the desired outcome without selecting a particular codec or transmission implementation. In the present example, the semantic SLA requires successful image delivery, sufficient reconstruction quality for downstream perception, and minimum communication cost among configurations satisfying the mandatory requirements.

In this experimental setup, sufficient visual quality is quantified as a mean PSNR of at least $30$~dB over the evaluated scenes, and successful delivery is required for every scene. Communication cost is represented by the channel bandwidth ratio (CBR), with lower CBR preferred once the delivery and quality constraints are satisfied. These values define an illustrative planning policy for the prototype rather than universal V2X service thresholds.

\begin{figure*}[t]
    \centering
    \includegraphics[width=0.96\textwidth]{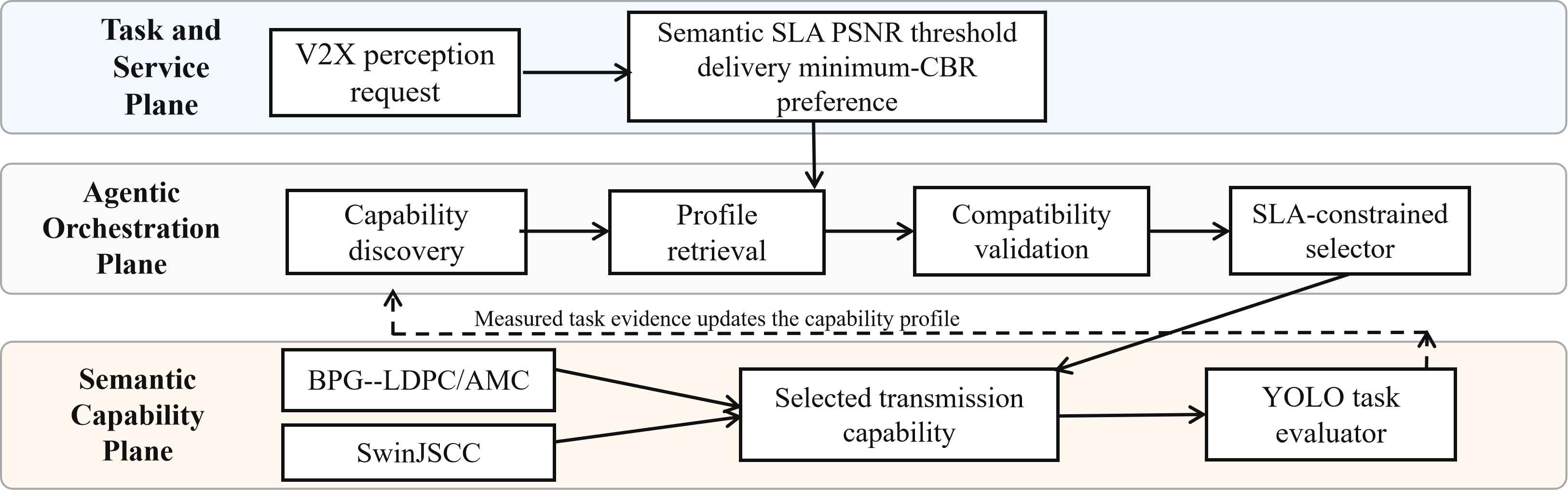}
\caption{Cross-plane workflow of the V2X case study. A perception request is translated into a semantic SLA specifying delivery, PSNR, and minimum-CBR requirements. The orchestration plane discovers candidate transmission capabilities, retrieves and validates their profiles, and selects the lowest-CBR feasible configuration. The selected BPG-LDPC or SwinJSCC configuration is executed and evaluated using YOLO, with task-level evidence fed back to update the capability profiles.}
    \label{fig:v2x_cross_plane}
\end{figure*}

As shown in Fig.~\ref{fig:v2x_cross_plane}, the orchestration plane retrieves feasible transmission profiles and presents their operating conditions and measured performance to the agent planner. Guided by the semantic SLA, the planner selects a transmission capability and its operating configuration. Before execution, deterministic guards verify model pairing, channel-use feasibility, delivery status, and compliance with the SLA. This organization separates open planning from safe execution. The planning space can expand as new capabilities and profiles become available, whereas compatibility and policy validation provide a stable execution boundary. The current minimum-CBR policy is one reproducible realization of this more general planning interface.

\subsection{Profiled Transmission Capabilities}

Two heterogeneous image-transmission capabilities are registered in the planning space. The first is a pretrained SwinJSCC model, fine-tuned on vehicle driving scenarios and supporting simultaneous SNR and rate adaptation~\cite {yang2024swinjscc}. The second is a separated digital capability consisting of BPG source coding, 5G NR LDPC channel coding, Gray-labelled modulation, and complex AWGN transmission. For each SNR-CBR point, a single modulation and coding scheme is selected across all source frames. The complete BPG bitstream is encoded using the lowest feasible QP under the available channel-use budget, followed by LDPC encoding and rate matching, QAM modulation, soft demapping, iterative decoding, and BPG reconstruction. Delivery is declared successful only when the recovered BPG payload matches the transmitted payload bit by bit.

The two capabilities need not share the same internal model structure. Instead, they expose a common evidence interface to the orchestration plane. Each operating profile records the supported input and output modalities, required encoder-decoder pairing, checkpoint or model version, SNR and CBR conditions, physical feasibility, delivery status, reconstruction quality, and available task-level observations. Explicit pairing information prevents invalid compositions, such as connecting an encoder to an unrelated decoder.

Sixteen $1024\times1024$ urban road images are used as source frames. The tested SNR values are $\{1,4,7,10,13\}$~dB, and the tested CBR values are $\{1/48,1/24,1/16,1/12,1/8\}$. Both capabilities use the same channel-use definition, with $\mathrm{CBR}=N_s/(3HW)$, where $N_s$ is the number of complex channel uses and $H$ and $W$ are the image height and width. The resulting measurements provide $400$ image-SNR-CBR observations for each transmission capability.

\begin{figure*}[t]
    \centering
    \includegraphics[width=0.96\textwidth]
    {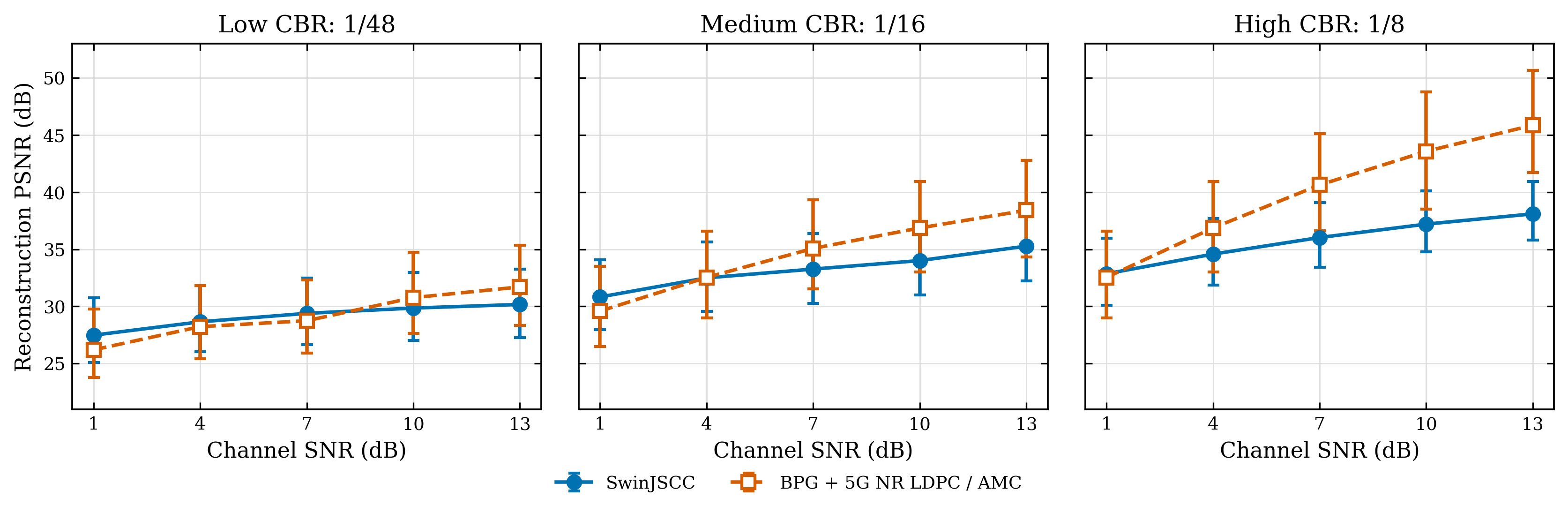}
    \caption{Measured operating profiles of the two candidate image-transmission capabilities. The panels show reconstruction PSNR versus channel SNR at CBRs of $1/48$, $1/16$, and $1/8$, with error bars indicating performance variation across the evaluated road scenes. The crossover between SwinJSCC and BPG+5G NR LDPC/AMC demonstrates that their relative performance depends jointly on channel quality and communication budget. These profiles provide condition-dependent evidence for SLA-based capability selection rather than a universal ranking of the two schemes.}

    \label{fig:capability_profiles}
\end{figure*}

Figure~\ref{fig:capability_profiles} illustrates the measured operating regions. SwinJSCC remains competitive in the low-SNR region, whereas the digital capability improves more rapidly with channel quality and provides higher reconstruction quality at sufficiently high SNR. The crossover depends on both SNR and CBR. These profiles are not intended to establish a universal codec ranking, but provide condition-dependent evidence that an agent planner can retrieve and compare when addressing a specific service request.

\subsection{SLA-Grounded Planning and Task Feedback}

For the experiment planning instance, the agent planner first retrieves all registered profiles associated with the requested image-transmission function. Compatibility and feasibility guards remove profiles with invalid model pairings, insufficient physical budgets, or unsuccessful delivery. Profiles whose mean PSNR is below the required threshold are also rejected. Among the remaining candidates, the planner selects the profile with the lowest CBR and uses measured PSNR as a tie-breaker.

This explicit policy makes the experiment decision reproducible, but it does not limit the architecture to a fixed selector. A different planning policy could consider task consistency, latency, energy consumption, privacy, execution history, or uncertainty in the registered measurements. New transmission or semantic-processing branches can also be added to the registry without changing the original task interface.

\begin{figure*}[t]
    \centering
    \includegraphics[width=0.96\textwidth]{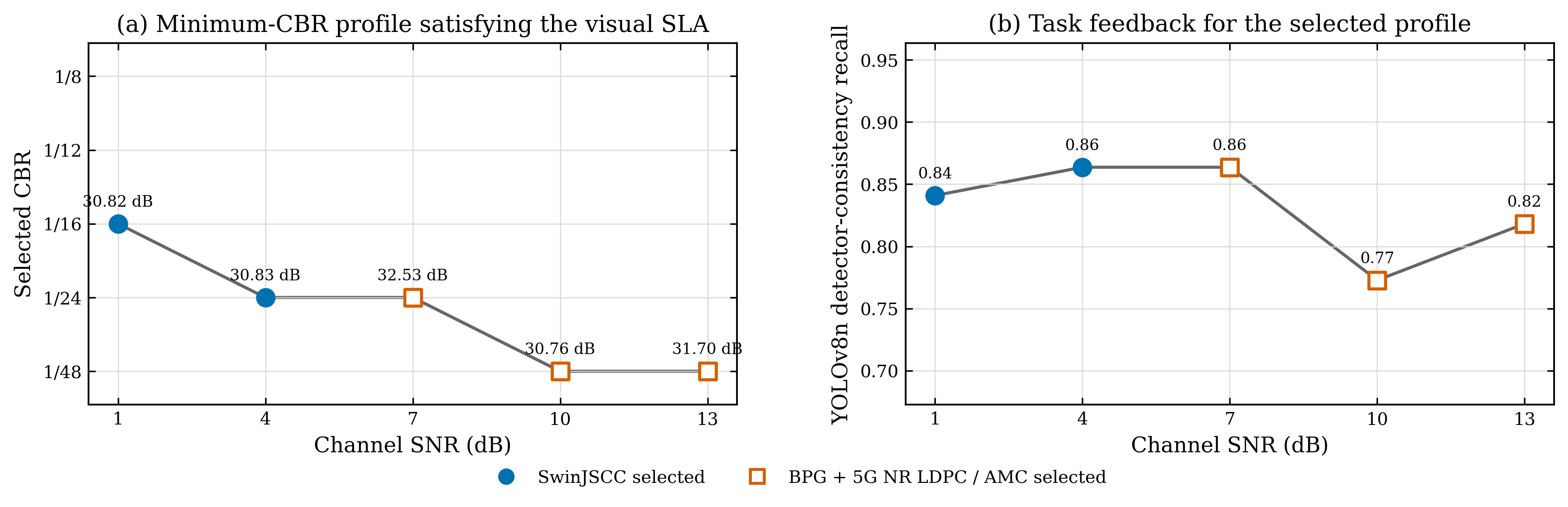}
\caption{SLA-constrained capability selection and the resulting task-level feedback. (a) At each channel SNR, the planner selects the lowest-CBR profile that satisfies successful delivery and the mean-PSNR threshold of $30$~dB. Marker types identify the selected transmission capability, and the annotations report its achieved mean PSNR. (b) YOLOv8n detector-consistency recall obtained from the same selected profiles. The recall is returned to the orchestration plane as task-level evidence but is not used by the current PSNR-based selection policy.}
    \label{fig:sla_selection_feedback}
\end{figure*}

The resulting plans are shown in Fig.~\ref{fig:sla_selection_feedback}. At $1$~dB, SwinJSCC is selected at $\mathrm{CBR}=1/16$, providing a mean PSNR of $30.82$~dB. At $4$~dB, SwinJSCC remains selected at $\mathrm{CBR}=1/24$, with a mean PSNR of $30.83$~dB. At $7$~dB, the planner switches to the digital capability at $\mathrm{CBR}=1/24$, achieving $32.53$~dB. At $10$ and $13$~dB, the digital capability satisfies the SLA at $\mathrm{CBR}=1/48$, with mean PSNR values of $30.76$ and $31.70$~dB, respectively.

The selections illustrate that a functional label such as ``image transmission'' is insufficient for capability planning. The two candidates provide the same high-level service but differ in quality and communication cost across operating conditions. The selected capability therefore changes with the channel condition even though the semantic SLA remains unchanged.

A fixed YOLOv8n detector~\cite{sapkota2025ultralytics} is registered as a task-evaluation capability. It is applied to the original and reconstructed images, with detections from the original image used as a pseudo-reference. A reconstructed detection is regarded as retained when it has the same task-level class and an intersection-over-union of at least $0.5$ with a reference box. Detector-consistency recall is then returned to the orchestration plane as task-level evidence. The detector-consistency recalls of the selected profiles are $0.84$, $0.86$, $0.86$, $0.77$, and $0.82$ at SNR values of $1$, $4$, $7$, $10$, and $13$~dB, respectively. The current planning policy uses PSNR and delivery status as mandatory constraints, while detector consistency is recorded as feedback rather than used directly for selection. In particular, the result at $10$~dB shows that satisfying the visual-quality requirement does not necessarily guarantee consistent task-level observations.

% \subsection{Implications and Scope}

% The case study demonstrates three properties of the proposed framework. First, a task request is separated from the implementation selected to satisfy it. Second, heterogeneous capabilities are compared through common, measurable profiles rather than internal model similarity. Third, agentic planning is combined with deterministic compatibility and policy guards, allowing the planning space to remain extensible without executing unsupported combinations. The measured reslut is reusable beyond the experiment selection policy. A transmission capability may be updated or replaced without modifying the service request, while a planner may adopt another objective without rewriting the internal transmission algorithms. Task evaluators can similarly be added or replaced to provide evidence aligned with a different semantic SLA.

% The current evaluation remains intentionally limited. It uses sixteen road scenes, one seeded channel realization for each image-profile pair, condition-specific light fine-tuning on forty  high-resolution images, and a restricted set of modulation and coding configurations. The digital capability implements a 5G NR LDPC-coded physical-layer link rather than a complete 5G NR protocol stack. The experiment selector is an explicit reproducible policy rather than an evaluation of unrestricted LLM planning. The results should therefore be interpreted as evidence of profile-grounded agentic planning and cross-plane interaction, rather than as a comprehensive V2X benchmark or a complete autonomous orchestration system.

\section{Open Challenges and Future Research Directions}
\label{sec:challenges}

Although SIoE provides a framework for composing communication and AI capabilities according to task-level requirements, several challenges must be addressed before it can support large-scale, dynamic, and safety-critical services.
\begin{itemize}
\item 
\textbf{Semantic SLA formalization and task-grounded evaluation:}
A semantic SLA must translate application intent into constraints that can be evaluated during capability discovery, planning, and execution. This translation is challenging because semantic quality is task-dependent and cannot be adequately captured by a single communication or reconstruction metric. As illustrated in the V2X case study, PSNR quantifies visual fidelity but does not fully indicate whether task-relevant objects are preserved. Future work should integrate communication-, semantic-, and task-level evidence, distinguish mandatory constraints from optimization preferences, and account for uncertainty in capability performance. 
% Another open problem is how task feedback should revise an SLA or trigger replanning without producing unstable or overly frequent pipeline changes.
\item 
\textbf{Capability contracts, interoperability, and lifecycle management:}
Reliable composition requires more than matching software input and output formats. Capabilities may depend on specific latent spaces, model checkpoints, normalization procedures, semantic units, channel assumptions, or preprocessing steps. Machine-verifiable capability contracts are therefore needed to describe compatibility, operating regions, resource requirements, model pairings, provenance, and version constraints. These descriptions must remain valid as models are updated, migrated to different hardware, or subjected to environmental drift. A complete capability lifecycle should encompass registration, profiling, validation, deployment, monitoring, updating, rollback, and retirement. Furthermore, cross-provider interoperability and registry synchronization must be achieved without requiring providers to disclose unnecessary proprietary implementation details.
\item 
\textbf{Scalable agentic planning and distributed execution:}
An extensible capability ecosystem gives rise to a large, constrained planning space that encompasses task quality, compatibility, channel conditions, computing resources, latency, privacy, and execution history. Agentic planners can interpret objectives and propose alternative workflows, but unrestricted planning may generate invalid combinations or introduce excessive control latency. A practical architecture may combine high-level agent reasoning with deterministic compatibility and policy validation, prevalidated pipeline templates, and fast local controllers. Further research is required concerning uncertain or incomplete capability profiles, concurrent service requests, pipeline-switching costs, and the decision boundary between parameter adaptation and structural replanning. Moreover, capability placement across devices, edge servers, and cloud platforms must jointly consider communication cost, computing availability, privacy, and response deadlines.
\item 
\textbf{Trustworthy, verifiable, and auditable execution:}
Agent-controlled capability composition introduces risks associated with inaccurate profiles, compromised models, manipulated intermediate representations, and poisoned task feedback or execution memory. Consequently, SIoE requires capability authentication, provenance tracking, access control, sandboxed execution, integrity-protected records, and deterministic enforcement of safety and privacy policies. Critical services may additionally require independent evaluators, redundant capabilities, or conservative fallback pipelines to verify task outcomes. A key research topic is how to maintain the flexibility of agentic planning while ensuring that every executed pipeline remains explainable, reproducible, and attributable to validated capabilities and policies.
\end{itemize}
\section{Conclusion}
\label{sec:conclusion}

This article proposes the SIoE, a composable architecture for coordinating heterogeneous communication and AI capabilities in alignment with task-level requirements. The three-plane framework separates semantic SLA specification, agentic planning and validation, and capability execution across distributed device, edge, and cloud resources. A lightweight V2X case study illustrates how profiled transmission capabilities can be retrieved, compared, and selected subject to explicit quality, delivery, and communication-cost constraints, while task-level measurements serve as feedback for subsequent planning. Although the current prototype employs a limited capability set and a reproducible planning policy, it demonstrates that communication and AI functions can be composed without embedding each service objective into a fixed end-to-end pipeline. Future work will focus on task-grounded SLA formalization, interoperable capability contracts, scalable agentic planning, and trustworthy execution.

\bibliographystyle{IEEEtran} 
\bibliography{ref} 

\end{document}